\documentclass[preprint,showpacs,aps]{revtex4}

\usepackage{graphicx}
\usepackage{dcolumn}
\usepackage{bm}

\begin{document}


\title{Fundamental modes of a trapped probe photon in optical fibers conveying periodic pulse trains}
\author{Alain~M.~Dikand\'e\footnote{Electronic address: adikande@ictp.it}}
\affiliation{Laboratory of Research on Advanced Materials and Nonlinear Sciences (LaRAMaNS), Department of Physics, Faculty of Science, University of Buea, P.O. Box 63, Buea, Cameroon}

\date{\today}
\begin{abstract}
Wave modes induced by cross-phase reshaping of a probe photon in the guiding structure of a periodic train of temporal pulses, are investigated theoretically with emphasis on exact solutions to the wave equation for the probe. The study has direct connection with recent advances on the issue of light control by light, the focus being on the trapping of a low-power probe by a temporal sequence of periodically matched high-power pulses of a dispersion-managed optical fiber. The problem is formulated in terms of the nonlinear optical fiber equation with averaged dispersion, coupled to a linear equation for the probe including a cross-phase modulation term. Shape-preserving modes which are robust against the dispersion are shown to be induced in the probe, they form a family of mutually orthogonal solitons the characteristic features of which are determined by the competition between the self-phase and cross-phase effects. Considering a specific context of this competition, the theory predicts two degenerate modes representing a train of bright signals and one mode which describes a train of dark signals. When the walk-off between the pump and probe is taken into consideration, these modes have finite-momentum envelopes and none of them is totally transparent vis-\`a-vis the optical pump soliton.
\end{abstract}

\maketitle
\section{\label{sec:level1}Introduction}
Optical solitons have been the subject of intense investigations in variable-dispersion media because of the great potential they offer for both fundamental research and technological applications in optical communications~\cite{haseg}. Particularly interesting are the techniques of dispersion management~\cite{haseg} and wavelength-division-multiplexing (WDM), the first stands for means to design optical fiber lines with a periodically varying dispersion where the peak power of the dispersion-managed (DM) soliton, as a steady pulse~\cite{kumar}, can be made larger so that the signal-to-noise ratio is improved. As for the second, the large gain bandwidth of modern amplifiers favors multiplexing of several DM pulse solitons onto a single fiber link using WDM. DM solitons have demonstrated evident potentials for WDM and are thus a very attractive choice as robust transmission mode in all-optical, long-distance transmission lines. \\
One promising application which exploits the stability of optical solitons is the gating communication process~\cite{manas1,manas2,fuente,darren1,ottav,darren}, where DM pulses are expected to provide a robust bench for optical beam traps and reshaping. A key factor making such a process feasible is the possibility to vary the separation between multiplexed pulses due to their mutual interactions via the DM technique~\cite{menyuk1}, such that a sequence of mutually interacting DM pulses can imprint a periodic structure~\cite{torner,darren1} within the fiber acting like a lattice of optical waveguides (i.e. traps). Since some pioneer works~\cite{manas3} reporting a possible reshaping of probe photons by means of cross-phase modulation, in recent years a large amount of experiments has been performed on various materials and gives evidence that solitons are prime candidates for controlling light by light~\cite{fuente1,lan1,hang,ken}. Quite interesting, via this control it is possible to trap low-power signals which are essentially dispersives (i.e. linear) with high-power pulses that are robust against the dispersion. In this last respect optical-soliton-guided photon implementations can be very beneficial in the storage, transport and routing of qubits~\cite{shapiro1,shapiro2}, where the reshaped photon beams can acquire solitonic properties (which we refer to as photosolitons) thus increasing bit rates and propagation distances. In particular the soliton-guiding implementation offers the possibility to design self-reconfigurable waveguide circuits which are able to convey shape-preserving photonic waves of wavelengths equal or different from the pump solitons, as envisaged e.g. in all-optical switching~\cite{bermel} and beam-steering waveguide devices.\\
Here we address the problem from the standpoint of fundamental physics, considering a photon trapped in the waveguiding structure of  a periodic sequence of temporal pulses delivered by an optical fiber. As this issue is of great current interest, it is of primary importance to highlight salient features of the system predicted by an exact treatment of the theoretical models suggested for this phenomenon~\cite{manas1,manas2,darren,huang,kanna}.
\section{\label{sec:level2}The model and fundamental soliton solutions of the probe equation}
\begin{figure}\centering
\begin{minipage}[c]{0.45\textwidth}
\includegraphics[width=7.cm]{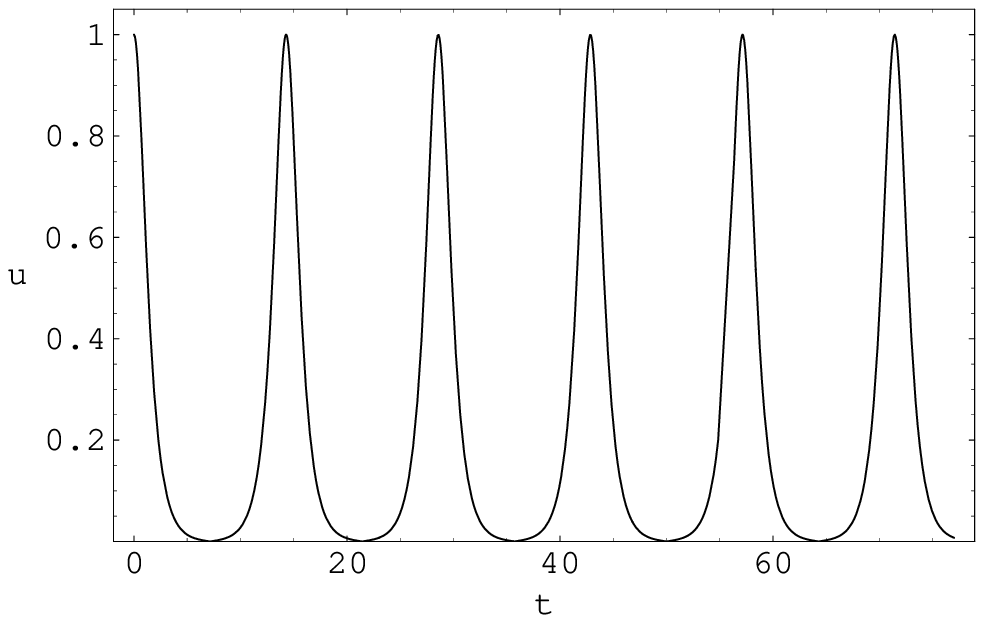}
\end{minipage}\vskip 0.3truecm
\begin{minipage}[c]{0.45\textwidth}
\includegraphics[width=7.cm]{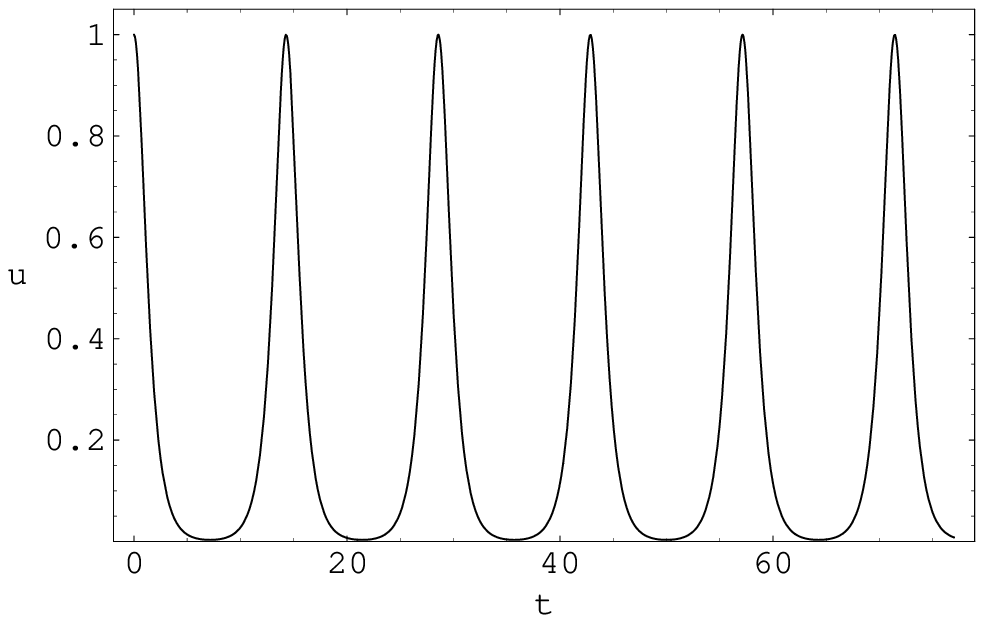}
\end{minipage}\vskip 0.3truecm
\begin{minipage}[c]{0.45\textwidth}
\includegraphics[width=7.cm]{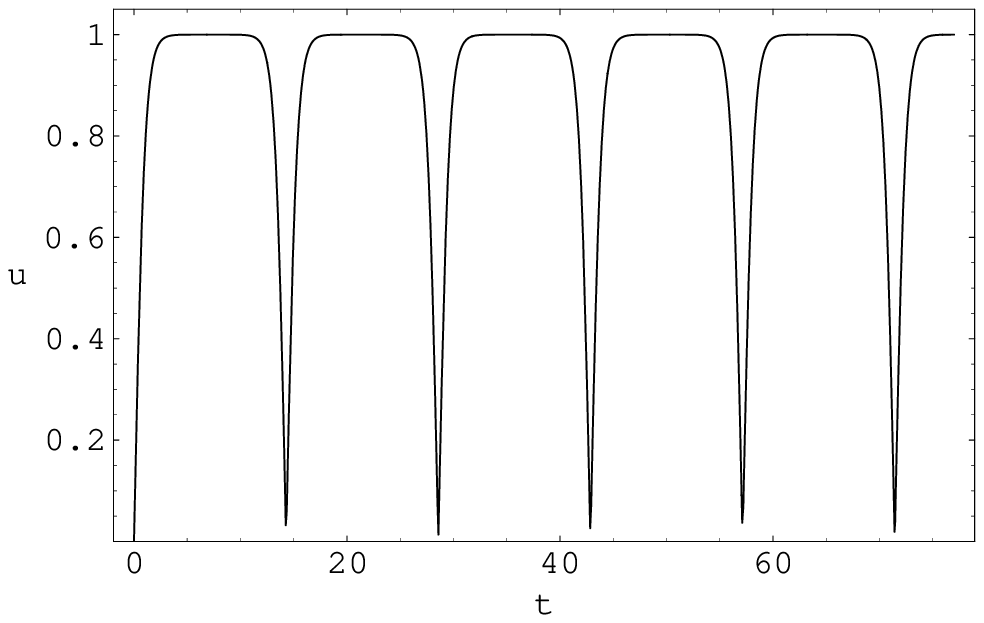}
\end{minipage}
\caption{\label{fig:one} Temporal profiles of the intensities $\vert A_k(0, t)\vert$ of the $cn$ (top), $dn$ (middle) and $sn$ (bottom) photosoliton modes for the value $\kappa=0.95$ of the Jacobi elliptic modulus.}
\end{figure}

The generic wave equations for the pump-probe system are represented by the coupled set~\cite{manas1,manas2,fuente,darren}: 
\begin{eqnarray}
i\frac{\partial q}{\partial z} &-& \frac{1}{2}D(z)\frac{\partial^2 q}{\partial t^2} + \xi\vert q \vert^2\,q= 0,
\label{a1a} \\
i\frac{\partial A}{\partial z} &-& \frac{1}{2}D_p(z)\frac{\partial^2 A}{\partial t^2} + i\lambda\frac{\partial A}{\partial t} + \beta\vert q \vert^2\,A =0,
\label{a1b}
\end{eqnarray}
where~(\ref{a1a}) is the variable-dispersion optical fiber equation~\cite{torner} and~(\ref{a1b}) is the equation for the photon probe~\cite{manas1,manas2,fuente,darren}. The quantities $q$ and $A$ are envelopes of the pump and probe respectively, $D(z)$ and $D_p(z)$ are their respective group velocity dispersions and $\xi$ is the nonlinear coupling coefficient of the pump source. $\lambda$ in the second equation is the walk-off between the probe and optical pump while $\beta$ measures the strength of the cross-phase interaction between the pump and probe. \\
As we are interested in optical signals generated in the anomalous regime of the group-velocity dispersions, it should be stressed that equation~(\ref{a1a}) is not integrable and so provides no exact analytical solution~\cite{lakob1}. For this last reason, variational approaches have been suggested, such as the collective-variable method~\cite{tmn1} in which a known trial function is used to describe analytically the pulse propagation along the fiber. In particular, the variational method has been recently employed~\cite{malomeda,Russell,Bao,Kockaert} to derive the evolution equations for the amplitude, width and chirp of DM solitons considering the lowest modes of the Hermite-Gaussian profile~\cite{kaup1}. However, Gaussian pulses are actually not robust enough to retain a full solitonic shape~\cite{menyuk2,turitsyn} and hence represent very crude approximations of the exact solution to the nonlinear optical fiber equation~(\ref{a1a}). The recognition of this shortcoming recently motivated new considerations for more appropriate variational pulses of the variable-dispersion nonlinear fiber equation, based on seeking for the exact solution to this equation but with a constant dispersion. The last picture is known as the averaged-dispersion approach and furnishes an efficient way of managing the fiber variable dispersion without sacrifying the fundamental feature of the signal, related principally to its solitonic properties~\cite{nak,nak1,nak2,dika}. \\ 
Here we shall follow this last picture, averaging the wave equations~(\ref{a1a})-(\ref{a1b}) over one map of the periodic dispersion in order to find exact expressions of the DM pulse solitons for the optical pump, and in turn for the probe. Setting $<D(z)>=-1$, we obtain the following steady-state DM soliton solution for the pump~\cite{torner,karta3,karta5}:
\begin{equation}
q(z, t)= \frac{Q}{\sqrt{\xi}}\, dn\left[Q\left(t-t_0 - \omega z\right),\kappa\right]e^{i\left[\omega t + \frac{\kappa^2\omega^2}{2} z + \frac{Q^2 - 2\omega^2}{2} z\right]}, \label{a2}
\end{equation}
in this expression $dn$ is the Jacobi elliptic function of modulus $\kappa$ while $Q$ is the amplitude and $\omega$ is the characteristic frequency of the soliton. The Jacobi elliptic function $dn$ is periodic in its arguments $(z, t)$ and for the solution~(\ref{a2}), the temporal period is $T= 2K(\kappa)/Q$ where $K(\kappa)$ is the elliptic integral of the first kind. \\ 
The many physical virtues of the Jacobi elliptic waves, connected with the problem of soliton transmission in nonlinear optical fibers, have been discussed at length~\cite{torner,nak,karta3,karta5}. Its main distinctive feature resides in the balance between the fiber dispersion and self-phase modulation (or nonlinearity) that guarantees the existence of such a shape preserving, well matched, yet interacting multiple pulses signal profile in the optical fiber. As for the question of which specific physics of the averaged-dispersion optical fiber is best described by such periodic structure, this issue has been addressed in ref~\cite{nak} from the standpoint of fundamental mathematical physics by solving the averaged-dispersion equation for the single optical fiber, with appropriate boundary conditions. Thus the author established that the $dn$ solution may be a suitable simplified representation of the state of inline time-division-multplexing of co-propagating pulses~\cite{nak}, a fact which can be emphasized by more elementary arguments exploiting Fourier series representations of the $dn$ function. Indeed, let us assume an optical fiber with an input signal consisting of a massive injection of single pulses at equal time interval $\tau_0$ in a multiplexer at the fiber entry $z_0$. Suppose the amplitude of the time-multiplexed pulses can readily be represented by the formal expression~\cite{menyuk1}:
\begin{equation}
\vert q_s(z_0, t)\vert= Q_n\sum_{n=1}^N{sech\,Q_n\left(t - t_n - \omega_n z_0\right)}, \label{a3}
\end{equation}
where $N$ is the number of injected pulses, $Q_n$ and $\omega_n$ are the amplitude and central frequency of the $n^{th}$ pulse respectively while $t_n=t_0 + n\tau_0$ is its initial temporal position. As pulses in the soliton lattice are equally separated, the multiplixing should be collisionless such that no frequency shift occurs. However, this is possible provided all pulses have same central frequency $\omega_0$ and equal amplitude $Q_0$. With these considerations, the sum over $n$ in~(\ref{a3}) becomes exact and the amplitude of the initial input signal after multiplexing, reduces exactly to a $dn$ function~\cite{hansen,magnus,malomed} of time coordinate. It is worthwhile underlining that this one-to-one equivalence between the $dn$ solution and the sum over the constituent individual pulses, can enhance substantially our theoretical ability to formulate the multiplexing of identical pulse signals. Namely it offers the possibility to express characteristic parameters of the $dn$ soliton as functions of parameters of individual pulses, hence permitting a better control of the stability of the $dn$-form pulse multiplex by judiciously choosing characteristic parameters of the constituent individual pulses. \\ 
Returning to the set of equations~(\ref{a1a})-(\ref{a1b}), we next look for shapes and characteristic parameters of the probe waves governed by equation~(\ref{a1b}). In this purpose, remark that since the nonlinearity in this equation is entirely related to the cross-phase term, any nonlinear wave that may be induced in the probe should be considered as a degenerate mode of the pump signal acting like an optical trap. To ensure full account of the time-multiplexing of the input pump at the fiber entry, we must then express the envelope of the probe as:
\begin{equation}
A_k(z, t)= u(t)\,e^{-i\left[k z - \phi(t)\right]}, \label{a4}
\end{equation}
where $u(t)$ is the core of the probe wave envelope, $k$ is the wavector associated with its spatial modulations and $\phi(t)$ is a characteristic temporal phase. Inserting~(\ref{a4}) in~(\ref{a1b}) and taking $D_p=-d$($d>0$) we obtain:      
\begin{equation}
-\frac{\partial^2 u}{\partial \tau^2} + \left[\frac{\beta}{\xi d} \kappa^2 sn^2(\tau) -  h(k) \right]u=0, \hskip 0.3truecm \phi_t= -\frac{\lambda}{d}, \label{a5}
\end{equation}
\begin{equation}
\tau= Q (t - t_0), \hskip 0.3truecm h(k)= \frac{\beta}{\xi d} - \frac{\epsilon}{Q^2 d}, \hskip 0.3truecm \epsilon= k - \frac{3\lambda^2}{2d}. \label{a5b}
\end{equation}
By setting 
\begin{equation}
\ell (\ell + 1)= \frac{\beta}{\xi d}, \label{a5c}
\end{equation}
equation~(\ref{a5}) becomes:
\begin{equation}
u_{\tau \tau} + \left[h(k) -\ell (\ell + 1)\kappa^2 sn^2(\tau)\right] u=0, \label{a6} 
\end{equation}
which is Lam\'e's equation~\cite{ars,dika1,dika2}. Note that when $\kappa=1$, this equation can be reduced to the Associated Legendre equation as obtained in~\cite{darren} in the case of a photon trapped by two interacting pulses which retain their individual shapes. \\
The Lam\'e equation possesses a rich spectrum with a great variety of eigenmodes~\cite{bill}, but the most interesting to us are those displaying a permanent profile in accordance with their solitonic properties. These modes are more exactly the boundstates of the Lam\'e equation, and because their formation via the cross-phase effect involves low cost to the pump in terms of momentum transfers they can be looked out like the groundstate modes of the spectrum of trapped states in the probe. The discrete states of Lam\'e's equation form a complete set of finite orthogonal modes which population depends on the integer quantum number $\ell$~\cite{dika1,dika2}. Quite remarkably, according to formula~(\ref{a5c}) the value of $\ell$ is determined by the competition between the self-phase-modulation effect responsible for the fiber nonlinearity and the cross-phase modulation effect on the probe exherted by the pump trap. We consider the lowest value of $\ell$, corresponding to the condition $\beta=2\xi d$ and for which $\ell=1$. This leads to the following family of modes:
\begin{equation}
u^{(1)}(\tau)= u_1 cn\left(\tau - \tau_0\right), \hskip 0.1truecm k^{(1)}=k_{\lambda} + Q^2d, \label{a7} 
\end{equation}
\begin{equation}
u^{(2)}(\tau)= u_2 dn\left(\tau - \tau_0\right), \hskip 0.1truecm k^{(2)}=k_{\lambda} + \left(2 - \kappa^2\right)Q^2d, \label{a8} 
\end{equation}
\begin{equation}
u^{(3)}(\tau)= u_3 sn\left(\tau - \tau_0\right), \hskip 0.1truecm k^{(3)}=k_{\lambda} + \left(1 - \kappa^2\right)Q^2d.  \label{a9}
\end{equation}
In addition to form a complete set of mutually orthogonal states, the three probe modes~(\ref{a7})-(\ref{a9}) equally display a number of interesting physical properties one most stricking being their shape preserving property. However, unlike the pump soliton which shape-preserving feature and periodic pulse matchings are entirely governed by the balance between the fiber dispersion and nonlinearity, for the probe the competition between the self-phase and cross-phase interactions is the governing factor of both the existence and stability of periodic structures in the probe. \\
The quantity $k_\lambda= 3\lambda^2/2d$ appearing in~(\ref{a7})-(\ref{a9}) introduces a finite low energy cut-off in characteristic momenta of the three modes, and reflects the walk-off preventing fully transparent(i.e. dispersionless) modes in the probe. Physically, this means it costs a finite amount of energy to the pump to create each of the three modes in the presence of the walk-off. \\
On figure~\ref{fig:one}, intensity profiles of the three fundamental modes are sketched for arbitrary values of their constant amplitudes $u_{i=1,2,3}$ and for the value $\kappa=0.95$ of the Jacobi elliptic modulus. 
It is quite noticeable that all of them represent nonlinear signals, but with distinct solitonic features. Indeed, if we proceed to an analysis of the three Jacobi elliptic functions in~(\ref{a7})-(\ref{a9}) via their Fourier series representations similar to~(\ref{a3}), we realize that the two first modes mimic two sequences of pulse signals with different characteristic momenta, while the third mode is equivalent to a periodic train of kink-shaped (i.e. dark) solitons. Moreover, the two first modes are of higher momenta as formula~(\ref{a7})-(\ref{a9}) suggest, and in addition represent two mutually degenerate states in the probe groundstate which energies and shapes coincide in the limit  $\kappa\rightarrow 1$. We note about this last observation that in the limit $\kappa\rightarrow 1$, both $cn$ and $dn$ transform to $sech$ (i.e. a single bright soliton mode) while $sn$ becomes $tanh$ (i.e. a single dark soliton mode). Also instructive to point out, in the absence of the walk-off the probe mode describing a dark soliton lattice has a vanishing momentum and hence is totally transparent vis-\`a-vis the optical bright soliton lattice constituting the pump. 
\section{\label{sec:level3}concluding remarks}
In summary, we carried out a systematic investigation of the low-energy modes of a probe photon trapped in the periodic sequence of temporal pulses delivered by an optical fiber with averaged dispersion. Our interest to these specific modes stems from their expected shape-preserving property which is desired to stabilize trapped states with long lifetime in the probe via the cross-phase modulation. We established that from the standpoint of fundamental physics, these fundamental modes are discrete states of the Lam\'e equation forming a complete set of mutually orthogonal photosoliton modes, which characteristic parameters are determined by the competition between the self-phase-modulation effect of the pump and the cross-phase-modulation effect of the optical pump on the probe photon. In particular, this competition governs the spectral broadening of the states associated with the photon trapping thus fixing the exact number of induced photosoliton modes. \\
In ref.~\cite{darren}, using the model equations~~(\ref{a1a})-(\ref{a1b}) the authors proposed an implementation describing the trapping of one photon by an optical soliton from a Kerr nonlinear medium. They shown that the equation governing modes induced in the probe by cross-phase modulation could be represented as the associated Legendre eigenvalue equation, and determined conditions for the existence of a mode with a robust shape profile in the case of a pulse soliton trap. They established numerically that because of the robustness of this probe carrying wave mode, the photon could be transferred between the captor soliton and another soliton with a minimum detrimental effect on the shape of the trapped photon mode, except possible energy losses from the phase shifts induced by collisions between the two interacting Kerr optical solitons. \\ 
It is nowadays widely established that the propagation of a photon packet can significant impact their emission and transfer, due to parametric recombinations and down conversions of the input photons in a nonlinear waveguiding structure. This manifests itself in several aspects of the emission and transfer processes, including the correlations of the matched modes and, depending on the phase-matching conditions, in the output spectrum which can be radically altered by multiple collisions between the guiding pulses. The degree of entanglement within a mutisoliton signal is thus highly dependent on the robustness of the multiplexing process, and in the case of matched soliton signals forming a waveguiding structure the photon transfer can become seriously empedded by processes like quantum phase noises, alteration of the waveguiding structure~\cite{darren1,darren} and so on if collisions between the guiding solitons are not controllable. In this last respect, the perfect matching of interacting pulses within the $dn$ soliton multiplex stands for a very powerful mean for overcoming several of these shortcomings, in particular the robustness of this periodic structure is a relevant minimizing factor of energy loss which is one of most important detrimental processes to be expected during the photon-mode transfer in an optical waveguide involving multiple interacting pulses.

\acknowledgments
I wish to thank T. C. Kofan\'e and A. B. Moubissi for enriching exchanges, and the Abdus Salam International Centre for Theoretical Physics (ICTP), Trieste Italy for support within the framework of the Regular Associateship scheme of the centre.

\end{document}